\documentclass[conference]{IEEEtran}

\usepackage{obfsSoK}

\ifCLASSINFOpdf
  \usepackage[pdftex]{graphicx}
  \DeclareGraphicsExtensions{.pdf,.jpeg,.png}
\else
\fi
\begin{document}

\title{Privacy engineering through obfuscation}

\author{\IEEEauthorblockN{Ero Balsa}
\IEEEauthorblockA{imec-COSIC, KU Leuven\\
ero.balsa@cornell.edu\textsuperscript{\textsection}}
}

\maketitle

\begingroup\renewcommand\thefootnote{\textsection}
\footnotetext{This paper was entirely written while the author was at KU Leuven.}
\endgroup

\begin{abstract}
	
	\emph{Obfuscation} in privacy engineering denotes a diverse set of data operations 
	aimed at reducing the privacy loss that users incur in by participating in digital systems.
	Obfuscation's domain of application is vast: privacy-preserving database analysis, 
	location-based privacy, private web search or privacy-friendly recommender systems
	are but a few examples of the contexts in which privacy engineers have resorted to obfuscation.
	Yet an understanding of the role that obfuscation, in general, 
	plays in the engineering of privacy has so far proved elusive. 
	Similarly, 
	we~lack a cohesive view of the wide array of privacy measures
	that assist the evaluation of obfuscation technologies.
    This paper contributes to closing these research gaps. 
    First, we~provide a general analysis framework 
	that brings together a multiplicity of obfuscation methods under the same analytical umbrella. 
	Second, we distinguish between mechanism-centred and attack-centred evaluation,
	making explicit a hierarchy of assumptions behind privacy measures
	that assists and demystifies obfuscation tools' evaluation.
	Finally, we examine the role that obfuscation technology plays in privacy engineering
	by introducing the concepts of \emph{personal} and \emph{public utility} 
	and distinguishing between utility-degrading and utility-preserving obfuscation.
	We observe that public utility requirements require us to resort to utility-degrading obfuscation to arbitrarily
	reduce privacy loss. Conversely, personal utility requirements do not, \emph{in theory}, impose such a privacy-utility trade-off,
	and we illustrate how to perform utility-preserving obfuscation through \emph{chaff}.

\end{abstract}

\IEEEpeerreviewmaketitle


\section{Introduction}

In computer security, \emph{obfuscation} denotes a family of techniques to achieve disparate security and privacy properties.
In software engineering and cryptography, obfuscation mainly denotes a set of transformations
to source or machine code that make it difficult for humans to understand and reverse-engineer software programs, 
e.g.~the program's purpose or internal values. 
Software engineers and cryptographers approach program obfuscation in radically different ways 
\mbox{---the}~former using \emph{security through obscurity} techniques, the latter favouring \emph{open} and \emph{provable}
security and formal properties--- however, program obfuscation still denotes a somewhat narrowly defined and concrete problem~\cite{kuzurin2007concept}.
Conversely, privacy engineers use the term \emph{obfuscation} to describe an array of methods and techniques 
that serve a disparate set of ends and purposes in various contexts such as web search privacy~\cite{obpws,gervais2014quantifying}, 
location privacy~\cite{ardagna2007location,duckham2005formal,shokri2011quantifying}, statistical disclosure control~\cite{machanavajjhala2006diversity,Sweeney02} or anonymous communications~\cite{berthold2002dummy,juarez2016toward}, 
among others, pursuing privacy properties specific to each of these contexts.

Privacy engineers perform obfuscation using a host of data operations such as randomisation, 
addition and suppression, generalisation, shuffling and swapping~\cite{aggarwalSurvey}.
However, even if privacy engineers use the same set of techniques across contexts, 
they do not always refer to them as \emph{obfuscation}.
Whereas location privacy researchers often refer to the operations they perform on location data as \emph{obfuscation}~\cite{duckham2005formal, shokri2012protecting,takbiri2017limits},
database privacy researchers rarely do so, favouring alternative terms such as \emph{perturbation} 
or more specific terminology such as \emph{randomisation} or \emph{generalisation},
even if the set of techniques both use is oftentimes the same or analogous.
Moreover, while researchers do work across contexts, bringing insights and expertise from one area to another~\cite{gervais2014quantifying,shokri2011quantifying,gervais2017quantifying}, 
a broader, more general bird's-eye view of obfuscation methods across contexts has so far been elusive;
terminology is often inconsistent and piecemeal, and the common underlying rationale 
and analogies between diverse obfuscation methods is not made explicit. 
This prevents a more general understanding of the role obfuscation plays 
\mbox{---regardless} of the particular form it takes and context in which it is applied--- in the design of privacy technologies.
When and why do we use obfuscation as opposed to other mechanisms in the engineering of privacy?
How does obfuscation complement or replace alternative protection mechanisms?
When \emph{must} we use obfuscation and when does it represent a suboptimal but useful, even needed, solution?
What trade-offs do obfuscation methods impose on other requirements such as utility or cost, and under which conditions? 
While this set of questions has been examined for particular contexts and scenarios, 
under a more or less narrow set of assumptions, 
a more holistic and general understanding of the role that obfuscation plays in privacy engineering is still missing.

\emph{Obfuscation tools} enable us to pursue a wide range of privacy properties under diverse conditions and assumptions, 
e.g.~from the limited set of assumptions that differentially-private mechanisms rely on
to attain independence from adversaries' background knowledge,
to the more concrete adversaries' specification that distortion-based measures of privacy require~\cite{shokri2015privacy}.
The design and evaluation of obfuscation tools requires that we define the privacy properties we seek and under which assumptions.
Then, to assess the extent to which obfuscation tools satisfy those privacy properties, 
we need to determine how to operationalise such properties through the right privacy measures.
Whereas debate in the literature as to which privacy measures one must choose to evaluate obfuscation technology abounds, 
such debate has been largely confined to narrow contexts and applications, 
be it web search privacy, location privacy or anonymous communications~\cite{gervais2014quantifying,andres2013geo,li2018measuring}.
The types of measures we can select to evaluate obfuscation solutions, 
the reason why we should choose one measure over another, 
as well as the underlying set of assumptions on which these measures depend,
these are questions that call for a cohesive, comprehensive understanding of the conditions and assumptions 
under which any particular measure suitably assists the evaluation of obfuscation technologies. 

Hence, in this paper we provide \emph{a bird's-eye view of obfuscation} tools in privacy engineering.
To that end, a first contribution of this paper is to bring obfuscation methods under a common analysis framework,
thus making explicit the relationship across obfuscation techniques and enabling 
a holistic study and understanding of the role these operations play in privacy engineering.\footnote{
We note that we do not intend to replace or force the use of the term obfuscation across disciplines. 
Rather, by referring to them as \emph{obfuscation}, we simply bring these analogous techniques under the same analytical umbrella.}

The second contribution is to propose an evaluation framework for obfuscation technologies \emph{in general},
abstracting away from the particularities of specific contexts and 
laying bare the set of assumptions that privacy measures depend on
and their impact on the design and evaluation of obfuscation~technology.
We propose two approaches to evaluating obfuscation tools: a mechanism-centred evaluation and an attack-centred evaluation.
This conceptualisation enables us to make explicit the underlying assumptions and level of generality of
different privacy measures, as well as to examine the role these measures play
in the design and evaluation of obfuscation technologies, 
e.g.~when it makes sense to seek \emph{independence from the adversary's knowledge}
as opposed to when we need to focus on one adversary with a particular instance of background knowledge and attack strategy.

The third contribution is to equip the analytical framework for obfuscation we propose in this paper
with a set of conceptual tools that enable us to examine the role of obfuscation in privacy engineering. 
More specifically, we introduce the concepts of \emph{personal} and \emph{public utility},
and distinguish between two types of obfuscation: \gls{udo} and \gls{upo}. 
Through this conceptual framework, we observe that arbitrarily reducing privacy loss under public utility requirements
mandates the deployment of \gls{udo}: we \emph{must} resort to (utility-degrading) obfuscation,
no other privacy technology or method enables us to 
arbitrarily minimise the privacy losses that the production of public utility prompts.
Conversely, personal utility alone does not impose a trade-off between utility and privacy, 
meaning that deploying \gls{udo} to balance personal utility against privacy is suboptimal;
and yet, obfuscation still plays a key role in defending users against privacy-invasive
service providers that, be it for vested interests in data collection or the high costs involved
in setting up their systems according to stringent privacy-by-design principles, 
fail to protect users' privacy.

Lastly, the fourth contribution is to highlight the role of \emph{chaff}, namely, the generation of fake or dummy activity, 
as a means to perform \gls{upo}. We examine the conditions that enable or hinder the deployment 
of this obfuscation method and highlight that escaping the trade-off between utility and privacy
invokes an alternative trade-off between privacy and~cost.


\section{An abstract model of obfuscation tools}
\sectionmark{Abstract model}
\label{s:model}

An individual uses an online service that offers a set of functionalities~$\{\fui\}$,
e.g.~sending messages to friends, browsing the Internet, posting on a blog or searching the web.
We refer to this individual as the \emph{user} and abstract from the particular type of service;
the user performs service requests~$\real_i$
and obtains responses~${\ans_i = \fui(\real_i)}$
so that~$\fu : \Reals \rightarrow \Ans  $ with~$\Reals$ the universe 
of user requests and~$\Ans$ the universe of service~responses. 
A~request~$\real$ denotes not only the service the user solicits 
but also input data that the execution of the service requires,
e.g.~the message a user sends to a friend on an instant messaging app 
or the URL of the website she wishes to access from a set of search results.
We~more generally refer to~$\real$ as the user \emph{input} and to~$\ans$ as the system \emph{output}.
We~denote a \emph{sequence} of user inputs as~$\rseq = [\real_1,\real_2,\ldots,\real_n]$ and a sequence of responses as~$\aseq = [\ans_1, \ans_2, \ldots, \ans_n]$.
We~describe~$\rseq$ through a set of features or random variables $\rfeat$;
using superscripts $j$ to denote a particular feature or set of features, e.g.~timing, 
content, size of individual actions $\real$, or any other relationships~therein.
Moreover, the user's input~$\rseq$ feeds a set of \emph{public utility functions}~$\sui$ 
with~${\su: \Rseq \rightarrow \Sans}$, where~$\Sans$ is the universe of public utility function outcomes $\sans_i$,
e.g.~film recommendations or traffic-aware driving directions, 
based on a user's film reviews and position on the road while driving, respectively, 
that other (untrusted and potentially adversarial) users of the system can benefit from.

An adversary is interested in the input sequence~$\rseq$
to obtain some information $\prof = \fa(\rseq)$, 
where~$\prof$ represents e.g.~the type or frequency of actions a user performs, 
or any other information that one may infer from the user's input. 
Function ${\fa : \Rseq \rightarrow \Prof}$ thus maps the universe of input sequences $\Rseq = \{\rseq\}$  to the universe $\Prof = \{\prof\}$.
Allowing the adversary to obtain~$\prof$ invades the user's privacy.
Depending on its capabilities, the adversary may be able to obtain~$\rseq$ directly or indirectly,
using adversarial function~$\ps_\su$ to extract information about $\rseq$ from public utility outcomes~$\sans$ that derive from~$\rseq$.

We account for the utility users derive and the privacy loss they incur in from disclosing $\rseq$ (either directly or through $\sans$).
The user obtains \emph{personal utility}~$\util_\fu$ from functionalities~$\fui$ so that~${\util_\fu : \Rseq \rightarrow \mathbb{R}^{+}}$.
The user incurs privacy loss~$\priv$ so that~${\priv : \Rseq \rightarrow \mathbb{R}^{+}}$. 
%
Moreover, the user also ``produces'' \emph{public utility} by contributing to~$\sans$
so that~${\util_\su :  \Rseq \rightarrow \mathbb{R}^{+}}$.
\emph{Public utility} refers to utility a user produces for everyone, 
including people she does not trust, i.e.~(potential) adversaries.
We refer to users that produce public utility as \emph{utility producers}.
Consider a service like Google Maps, which provides functionalities~$\fui$ 
such as route planning and navigation based on real-time traffic conditions.
To plan and navigate a route in real-time, a user needs to input her location, desired destination and chosen means of transport~($\real_i$). 
At the same time, Google relies on users' location and navigation data to calculate the time it takes to complete a route, 
obtain real-time traffic conditions or detect roadblocks, among other elements involved in route planning. 
Hence, users contribute with their data~$\rseq$ to improve the navigation service for everyone (including themselves). 
If we consider that Google and other Maps's users are non-trusted, potentially adversarial
\mbox{---i.e.}~they may leak or abuse the information that other users provide while using the service---
then, we may refer to the utility these non-trusted entities obtain from~$\rseq$ as public utility.

Conversely, \emph{personal utility} refers to the utility an individual obtains from~$\fui$ 
exclusively for herself and for others she trusts. 
Consider a Google Maps user who does not want Google to leverage her data to provide utility to others, 
i.e.~she wishes to obtain route recommendations for herself
without letting Google use her data to improve recommendations for everyone.
In this case, the user exclusively seeks \emph{personal utility}, i.e.~she is a \emph{utility consumer}.
Another example: a user that sends a private message to a friend and intends that information to be
of utility to no one besides herself and her friend, whom she trusts. 
In short, users may exchange and reveal data among a trusted group of relatives, friends or coworkers. 
These exchanges do not represent public utility; 
rather, the utility users derive from disclosing~$\rseq$ to trusted peers is part of their personal utility.
Public utility exclusively derives from data disclosures to untrusted parties, e.g.~that the general public can benefit from.
Whereas this decision may seem arbitrary,\footnote{
	Deciding which peers or entities a user trusts is in fact entirely	at the user's or system designers' own discretion.
}
it enables us to establish a crisp separation between the disclosure requirements that 
either type of utility imposes and the role of obfuscation in privacy engineering, as we discuss in Sect.~\ref{s:upoUdoNmore}.

\begin{figure}
	\centering
	\includegraphics[width=\columnwidth]{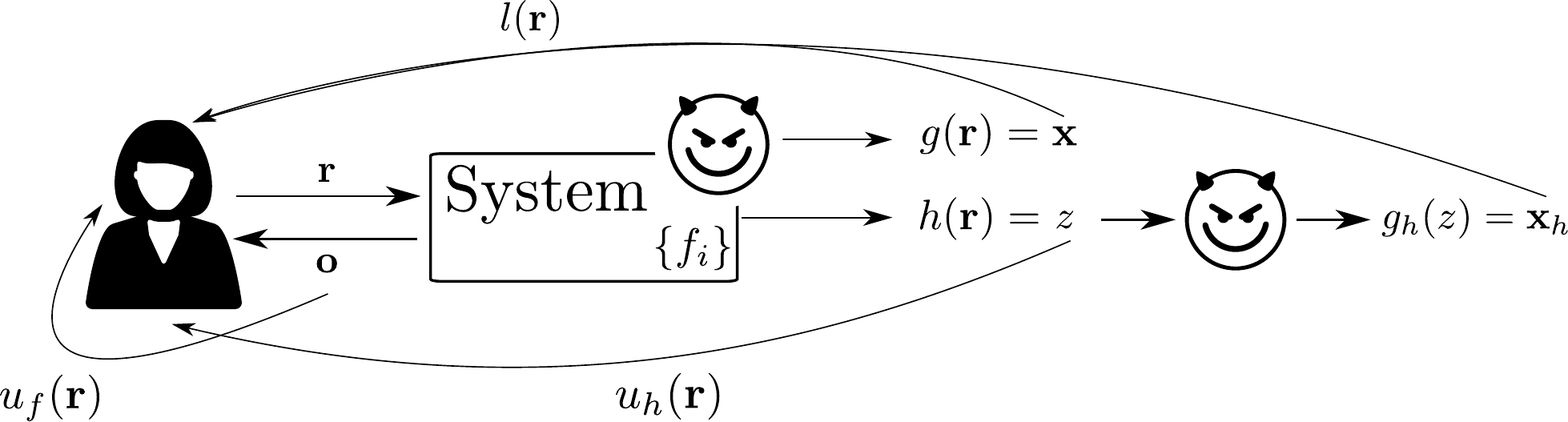}
	\caption[System and threat model]{System and threat model with utility and privacy loss flows.}
	\label{f:model}
\end{figure}

Figure~\ref{f:model} depicts the system and adversary model we have introduced in this section:
a~user submits to the system an input sequence~$\rseq = [\real_1, \real_2, \ldots, \real_n]$ 
and retrieves a sequence of outputs~$\aseq = [\ans_1, \ans_2, \dots, \ans_n]$, where $\ans_i=\fui(\real_i)$.
As a result, the user derives some \emph{personal utility}~$\util_{\fu}(\rseq)$ for herself.
Moreover, the system feeds the user's input~$\rseq$ to a public utility function~$\su$, releasing a public utility outcome~$\sans$ 
that produces some \emph{public utility}~$\util_{\su}(\rseq)$ for \emph{everyone}, including the user.
At the same time, by submitting~$\rseq$ to the system, the user incurs some privacy loss~$\priv(\rseq)$ due to 
two archetypal adversaries: one the one hand, advesaries that have access to the system itself, capturing input sequence~$\rseq$ 
and obtaining some information~$\prof = \ps(\rseq)$;
on the other hand, adversaries that exploit the public utility outcome~$\sans$ 
to some obtain information~$\prof_\su = \ps_{\su}(\sans)$ about the~user.
Table~\ref{tab:onObfsNotation} summarises the notation we use throughout this paper. 


\begin{table}[h!]
\centering
  {\renewcommand{\arraystretch}{1.5}
  \scriptsize
  \begin{tabular}{c p{0.33\columnwidth} c c p{0.33\columnwidth}}
  \multicolumn{2}{l}{Symbol \hspace{1cm} Meaning}  & & \multicolumn{2}{l}{Symbol \hspace{1cm} Meaning}\\
  \toprule
  $\real$ & User input/request & & $\rseq$ & Sequence $[\real_1, \real_2, \ldots, \real_n]$ \\
  $\Reals$ & Universe of user inputs & & $\Rseq$ & Universe sequences~$\rseq$  \\
  $\rseqrv$ & Random variable over $\rseq$ & & $\rfeat$ & Input sequence feature  \\ \rule{0pt}{5ex} 
  $\ans$ & Output/response, $\ans=\fu(\real)$ & & $\aseq$ & Sequence $[\ans_1, \ans_2, \ldots, \ans_n]$ \\
  $\Ans$ & Universe of outputs~$\ans$ & & $\Aseq $ & Universe of sequences~$\aseq$ \\  \rule{0pt}{5ex} 
  $\fu$ & Personal functionality & & $\util_\fu$ & Personal utility \\
  $\fa$ & Adversarial function & &  $\priv$ & Privacy loss function \\
  $\su$ & Public utility function & & $\putil$ & Public utility \\ \rule{0pt}{5ex} 
  $\prof$ & Adversarial  function's outcome, ${\prof = \fa(\rseq)}$ & & $\Prof$ & Universe of adversarial outcomes \\  
  $\profV$ & Random variable over $\prof$  & &   $\filseq$ & \emph{Deobfuscated} user input \\ \rule{0pt}{5ex} 
  $\sans$ & Public utility outcome & & $\Sans$ & Universe of~$\sans$ \\  \rule{0pt}{5ex} 
  $\rufeat$ & Feature contributes to utility & &   $\rnufeat$ & Feature \emph{does not} contribute to utility \\
  $\rlfeat$ & Feature contributes to privacy loss & & $\rnlfeat$ & Feature \emph{does not} contribute to privacy loss \\\rule{0pt}{5ex} 
  $\osktor$ & Obfuscation function & & $\mi$ & Mutual information \\
  $\gain$ & Information gain & & $\eee$ & Expected estimation error \\ \rule{0pt}{5ex} 
   $\eseq$ & Obfuscated sequence & & $\Eseq$ & Universe sequences~$\eseq$  \\
 $\eseqrv$ & Random variable over $\eseq$ & & $\obfsans$ & Obfuscated public outcome, $\obfsans = \osktor(\sans)$ \\
 $\dseq$ & Sequence of \emph{chaff} \\
 \bottomrule  \rule{0pt}{3ex} 
  \end{tabular}
  }
\caption[Notation]{Notation}
\label{tab:onObfsNotation}
\end{table}


\paragraph*{Obfuscation tools}
\label{obfTools}

An \emph{obfuscation tool} (\glsunset{ot}\gls{ot}) modifies~$\rseq$ or~$\sans$
with the aim of limiting or preventing privacy loss.
To that end, an obfuscation mechanism~$\osktor$ modifies~$\rseq$ or~$\sans$
to produce an obfuscated sequence ${\eseq = \osktor(\rseq)}$ or obfuscated output~$\obfsans = \osktor(\sans)$, respectively.
\Glspl{ot} may rely on different forms of obfuscation, such as the generation of chaff 
(i.e.~fake user activity) or, conversely, the suppression or discarding of user inputs~\cite{Sweeney02,ebalsaPhdThesis};
\ots\ may also modify individual user inputs~$\real_i$ into $\ev_i$ 
by generalising or adding random noise to \mbox{their~values~\cite{Sweeney02,aggarwal2008general,bertino2008survey}}.

As a result, the adversary observes an obfuscated input~$\eseq$ or obfuscated output $\obfsans$
that should ideally leak no or less information about~$\prof$, so that $\priv(\eseq) < \priv(\rseq)$ 
(respectively~${\priv(\obfsans) < \priv(\sans)}$).
However, obfuscation impacts utility too, with personal and public utility becoming $\util_{\fu}(\eseq)$ 
and $\util_{\su}(\eseq)$, respectively.

\begin{figure}
	\centering
	\includegraphics[width=\columnwidth]{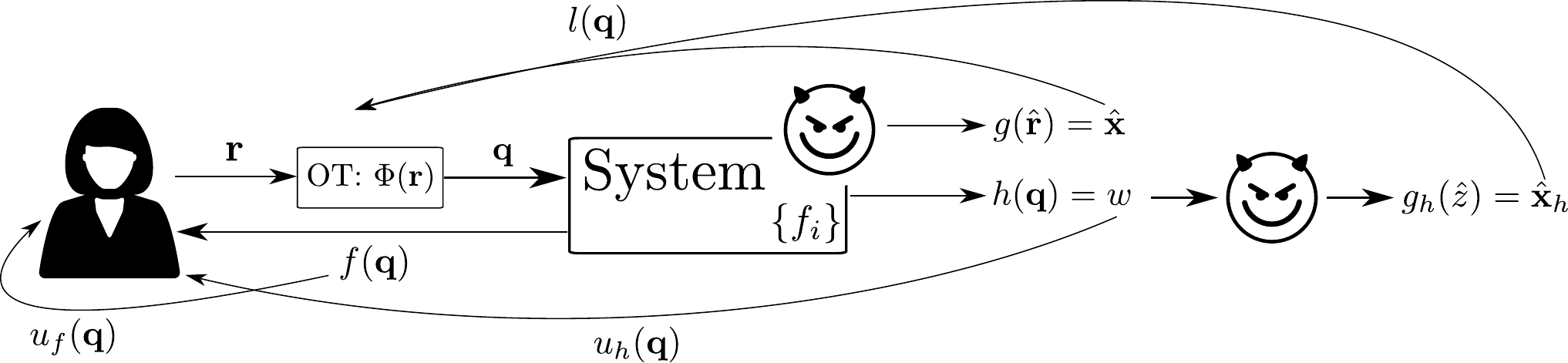}
	\caption{Process flow of obfuscation tools, user deployment.}
	\label{f:modelWithObfs}
\end{figure}

\begin{figure}
	\centering
	\includegraphics[width=\columnwidth]{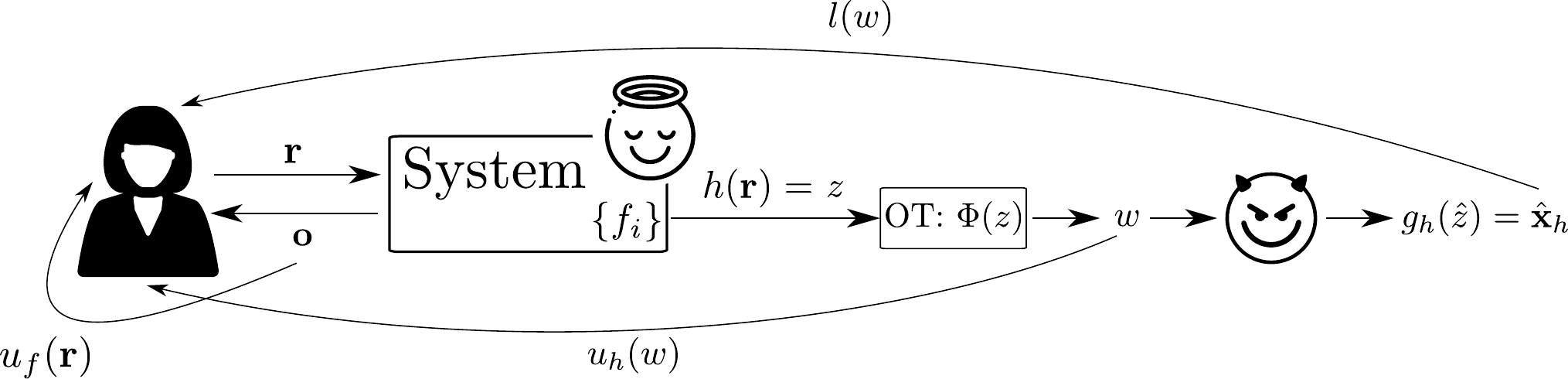}
	\caption{Process flow of obfuscation tools, service provider deployment.}
	\label{f:modelWithObfs2}
\end{figure}

\begin{figure}
	\centering
	\includegraphics[width=\columnwidth]{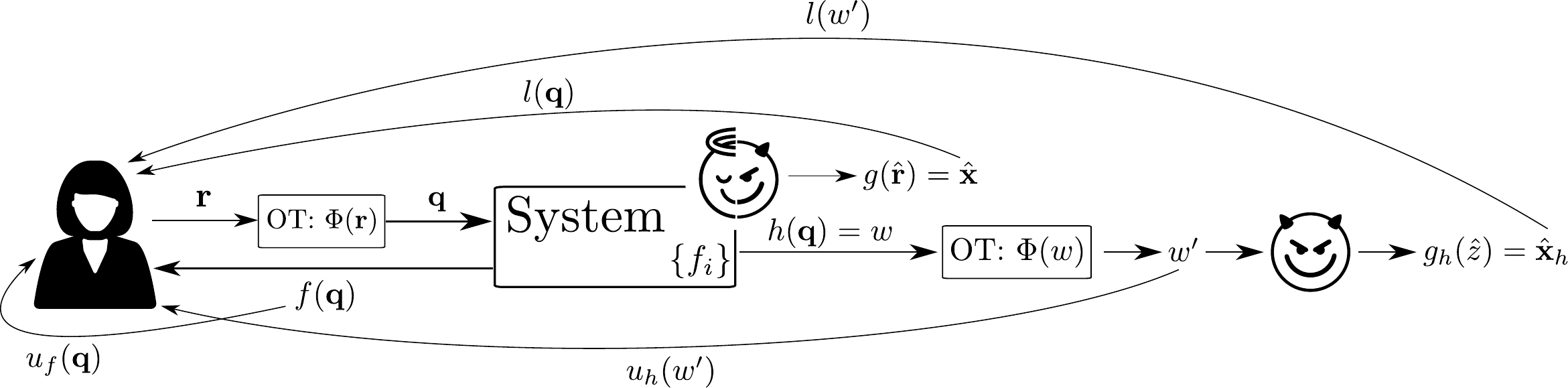}
	\caption{Deployment of \ots\ by both user and service provider.}
	\label{f:modelWithObfs3}
\end{figure}

Figure~\ref{f:modelWithObfs} depicts the deployment of an \ot\ by the user.
Alternatively, Fig.~\ref{f:modelWithObfs2} depicts the service provider obfuscating public utility outcomes, 
which implicitly assumes a trusted service provider that implements obfuscation on behalf of its users, 
e.g.~the trusted data curator that releases noisy statistics in the classic differential privacy model~\cite{dwork2014algorithmic}. 
Still, users may deem insufficient the level of protection that results from service provider obfuscation, 
leading to the deployment of obfuscation by both users and service providers, as Fig.~\ref{f:modelWithObfs3}~shows. 
For the sake of brevity, in this paper we focus on the first scenario, i.e.~user-deployment of obfuscation tools, 
as that offers protection against both types of adversaries so that
most of the observations and insights we provide in this paper implicitly extend to the latter two scenarios
(yet we~offer additional clarification where appropriate).

\paragraph*{Adversary model}
\label{s:advmodel}

As we have just mentioned, we consider two types of adversaries, 
depending on whether they have access to either~$\eseq$ or~$\obfsans$.
We focus on the first type of adversary but the assumptions and notation introduced below
trivially extrapolate to the second \emph{`external'} type of adversary. 

The adversary's goal is to obtain~${\prof = \ps(\rseq)}$.\footnote{
Analogously~${\prof_\su = \ps_\su(\sans)}$ for the `external' adversary.
}
The adversary knows that an~\ot\ is in place, i.e.~that ${\eseq = \osktor(\rseq)}$.
	The adversary has full knowledge about the \ot's design, i.e.~it can test and simulate runs of the~\ot\
	or perform an analytical evaluation of its obfuscation mechanism~$\osktor$, 
	e.g.~to determine the \ot's conditional probability distribution~$\cp{\eseqrv}{\rseqrv}$, 
	where $\rseqrv$ is a random variable (r.v.) that describes users' input sequences~$\rseq$
    and $\eseqrv$ the r.v.\ that describes obfuscated sequences~$\eseq$.
	We make no assumptions about the adversary's \emph{background knowledge}
	or \emph{auxiliary} information on users, which may relate to a particular individual,
	to an \emph{``average''} user or to the general population.

	Lastly, the adversary is \emph{strategic}, i.e.~it attempts to retrieve $\prof$ 
	\emph{``undoing''} obfuscation to the best of its abilities. 
	The adversary thus recovers an approximation $\filprof$ so that, ideally, $\filprof = \prof$.
	Note that Figures~\ref{f:modelWithObfs} through~\ref{f:modelWithObfs3} implicitly depict the adversary's attack
	whereby it obtains $\filprof$ and $\hat{\sans}$ from $\eseq$ and $\obfsans$, respectively.
	Still, the adversary is honest-but-curious (\glsunset{hbc}\gls{hbc}), 
	i.e.~it does not interfere with users' service requests ---regardless of its ability to do so---;
	it does not drop, delay or modify any (obfuscated) user actions.
	As an example of this kind of adversary, 
	we consider an \emph{honest-but-curious adversarial service provider} throughout this paper.


\section{Evaluating obfuscation tools}

\begin{figure}
	\centering
	\includegraphics[width=.85\columnwidth]{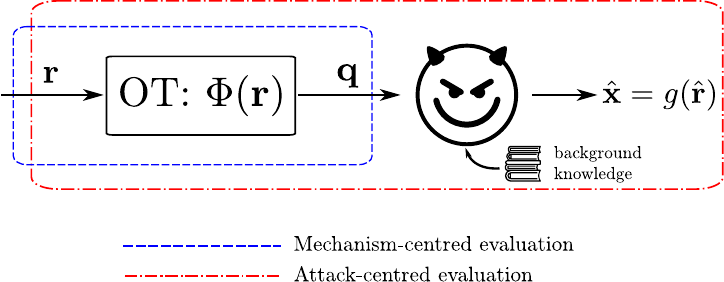}
	\caption[Mechanism-centred and attack-centred analyses]{Mechanism-centred and attack-centred evaluation.}
	\label{fig:cpoAnalyses}
\end{figure}

This section introduces a framework for the evaluation of \ots.
We unpack key assumptions underlying privacy measures for the design and evaluation of \ots. 
To that end, we distinguish between two approaches to measuring the level of privacy protection \ots\ afford: 
\emph{mechanism-centred} and \emph{attack-centred}. 
A mechanism-centred approach evaluates \ots\ with independence from a particular adversary 
or attack strategy, they focus on the relation between an \ot's inputs and outputs, 
rather than on the knowledge or exploits of a particular adversary;
conversely, attack-centred approaches focus on more detailed adversaries and attacks.
\Gls{mca} measures relate to the \ot's \emph{leakage}, 
this is, how much information \ots\ leak or disclose
regardless of the ability of an adversary to exploit that information. 
\Gls{aca} measures on the other hand relate to adversarial \emph{retrieval},
this is, the extent to which an adversary takes advantage of 
the information an \ot\ leaks to recover~$\prof$.
Figure~\ref{fig:cpoAnalyses} illustrates how 
\gls{mca} focuses on the relationship between an \ot's input $\rseq$
and output $\eseq$, abstracting away from a particular attack,
whereas \gls{aca} does take into account the adversary's knowledge and attack, 
capturing the combined, intertwined effect of obfuscation tool and adversarial attack,
i.e.~evaluating an \ot's ability to mitigate privacy loss through a measure of adversarial success.

\subsection{Mechanism-centred analysis}
\label{s:mca}

To illustrate the type of privacy measures that fall within mechanism-centred analysis, 
we review measures of \emph{indistinguishability} and measures of \emph{information leakage}. 


\subsubsection{Indistinguishability} 
\label{par:ind}

With its roots in cryptography, indistinguishability measures an \ot's ability to generate obfuscated inputs $\eseq$
so that any sequence~$\rseq$ could have led to~$\eseq$, rendering an adversary unable 
to determine the original input~$\rseq$ with higher than random chance, i.e.~based on the \emph{prior} probability $\pr{\rseq}$ alone.

Several works have proposed privacy definitions based on the notion of indistinguishability~\cite{hermans2011new,yao2006indistinguishability},
the most prominent to date being \emph{differential privacy}~(\glsunset{dp}\gls{dp})~\cite{dwork2017calibrating}.
Initially defined as $\e$-indistinguishability, \gls{dp} requires that the outcome $\eseq = \osktor(\rseq)$ is equally likely (within a multiplicative factor $e^\e$) 
if the input to the function were an input $\rseq'$,
for all pairs of inputs $\rseq$ and $\rseq'$ and all possible $\eseq \in \Eseq$.
An obfuscation tool thus ensures $\e$-indistinguishability
if the probability of any input $\rseq$ or $\rseq'$ leading to 
obfuscated input $\eseq$ is bounded by a multiplicative factor $e^\e$, namely:
\begin{equation}
	\label{eq:eldp}
	\sup_{\set \in \Eseq; \; \rseq, \rseq' \in \Rseq} 
	\frac{ \osktor( \set \given \rseq ) }{ \osktor(\set \given \rseq') } \leq e^\e
\end{equation}

Note that we do not specify that $\rseq$ and $\rseq'$ differ on at most in one record or action $\real$
as the classic differential privacy definition does.
In fact, since we focus on the user deploying an obfuscation tool, 
Eq.~\ref{eq:eldp} corresponds to a definition of \emph{local} differential privacy~\cite{kairouz2004}. 
Alternatively, a (trusted) curator releasing public outputs~$\sans$
may adopt a differentially-private obfuscation mechanism that satisfies:

 \begin{equation}
 \label{eq:dp}
 \sup_{\set \in \Sans; \; D, D' \in \Rseq^{N}} 
 \frac{ \osktor( \set \given D ) }{ \osktor(\set \given D') } \leq e^\e
 \end{equation}
 
 where $D$ and $D'$ represent two datasets that differ at most on the contribution of one individual, 
 e.g.~$D$ could be the data the service provider has collected on every user, $D=[\rseq_1,\rseq_2,\ldots,\rseq_N]$
 so that $D$ and $D'$ differ at most on the sequence of actions~$\rseq$ of any user.

Many alternatives and variations of $\e$-indistinguishability or~\gls{dp} exist, 
such as statistical distance (an \emph{additive} ---as opposed to \emph{multiplicative}--- measure of indistinguishability),
relaxations such as $(\e,\delta)$-differential privacy and $(\e,\dist,\delta)$-differential privacy,
the particular choice depending on the desired stringency of the measure~\cite{dwork2017calibrating,dwork2006our,mcsherry2009differentially,bassily2015local}.
Researchers have also proposed proposed variants for specific contexts, 
such as location privacy or anonymous communications~\cite{andres2013geo,backes2013anoa}.

The key point is that these indistinguishability measures do not depend on how the adversary attacks or what the adversary knows:
they measure privacy loss by focusing on the inputs and outputs of an \ot\ alone, 
thereby providing a measure of privacy independent from adversarial knowledge.

\subsubsection{Information leakage}
\label{s:infoLeak}

\begin{figure}
  \centering
  \includegraphics[width=.85\columnwidth]{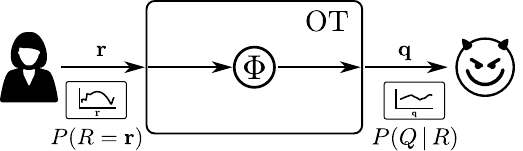}
  \caption[Protos as noisy channels]{An \ot\ as a noisy channel. Output sequences $\eseq$ result from the obfuscation of $\rseq$.}
  \label{f:noisyChannel}
\end{figure}

We may conceptualise \ots\ as communication channels through which the user
transmits a sequence of messages $\rseq$ to the adversary, 
who receives a \emph{noisy} sequence~$\eseq$, 
as Fig.~\ref{f:noisyChannel} shows.
Information theory enables us to measure the amount of information the channel (i.e.~the \ot) \emph{leaks},
i.e.~how much information $\eseq$ provides about $\rseq$.
If the channel is \emph{perfectly noisy}, $\eseq$~leaks no information about input~$\rseq$:
it is no easier for an adversary to determine $\prof$ after obtaining $\eseq$ than before, 
i.e.~with \emph{prior information} about $\prof$ alone.
Conversely, if the channel introduces no noise, 
obfuscated input~$\eseq$ univocally reveals the \ot's input $\rseq$ 
\mbox{---so}~the adversary can trivially compute $\prof$.

To measure the amount of information obfuscated sequence~$\eseq$ leaks about~$\rseq$, 
we compute the \emph{mutual information}~$\mi(\rseqrv;\eseqrv)$:

\begin{equation}
\label{eq:mi}
\begin{split}
\mi(\rseqrv;\eseqrv) 	& \equiv \sh(\rseqrv) - \ch{\rseqrv}{\eseqrv} = \\ & =  \sum_{\rseq \in \rseqrv} \sum_{\eseq \in \eseqrv} \pr{\rseq,\eseq} 
\log \frac{ \pr{\rseq, \eseq} }{ \pr{\rseq}\pr{\eseq} }
\end{split}
\end{equation}

\gls{mi} provides a measure of the channel's information leakage, 
but only \emph{for a particular probability distribution} $\pr{\rseqrv}$~\cite{chatzikokolakis2008anonymity,damgard1998statistical};
\gls{mi} says little about the performance of an~\ot\ in general, 
this is, for any probability distribution other than~$\pr{\rseqrv}$.
To determine the \ot's information leakage with more generality,
we consider mutual information between inputs and outputs to the \ot\ 
over \emph{all} possible input probability distributions,
namely, the channel's \emph{capacity}, which is the maximum information leakage
over all possible probability distributions of the input r.v.~$\rseqrv$.
Formally:

\begin{equation}
\label{eq:capacity}
\begin{split}
\cc = \sup_{\prsub{\rseqrv}{\rseq}} \mi(\rseqrv;\eseqrv) 
\end{split}
\end{equation}

Ideally, a perfect obfuscator has $\cc = 0$, i.e.~it leaks no information on no matter what input probability distribution.
Channel capacity thus provides a tight upper bound of an \ot's information leakage.\footnote{
Yet, as Chatzikokolakis et al.\  concede,  whenever we are able to accurately determine the probability distribution $\pr{\rseqrv}$,
mutual information provides a tighter measure of information leakage
as we do not need to consider input distributions 
under which the obfuscator performs worse~\cite{chatzikokolakis2008anonymity}.
}

Similarly to indistinguishability measures, there are alternative measures of information leakage,
such as min-leakage and min-capacity or generalised notions based on gain functions 
such as $g$-leakage, that provide slightly different measures of information leakage~\cite{m2012measuring,smith2011quantifying}.
Appendix~\ref{appendixMI} provides a discussion on alternative measures of information leakage
that do not rely on Shannon's entropy. 



\subsubsection*{Independence from adversarial knowledge}

Under certain assumptions, \glsreset{mca}\gls{mca} provides privacy guarantees 
\emph{independently from adversarial knowledge}. 
Differentially-private mechanism design represents a prominent example of 
attempting to enforce such a property.
The guarantee differential privacy offers ---namely, that the probability of a privacy breach will be more or less 
the same whether or not a user participates in the system--- holds, 
under certain assumptions about the data-generation process, regardless of what the adversary knows~\cite{dwork2017calibrating,kifer2011no}.
Hence, it represents a property of the mechanism itself, 
unrelated to what the adversary does or knows~\cite{dwork2014algorithmic}. 
Similarly, a channel's capacity measures the maximum amount of leaked information,
providing an upper bound on what the adversary may feasibly extract from the \ot. 
Information-theoretical measures do however depend on the mechanism's input data probability distribution;
capacity is not independent from the input data probability distribution per se, 
yet by considering \emph{all possible} probability distributions, 
it also becomes independent from the adversary's knowledge~\cite{cuff2016differential}. 

\subsubsection*{Indistinguishability or information leakage?}

Either type of privacy measure provides different types of guarantees. 
Multiplicative measures of indistinguishability ensure a minimum level of privacy for each particular user input;
capacity on the other hand provides an \emph{average} of the amount of information the \ot\ leaks,
but no bounds on the distinguishability between any two inputs: 
the \ot\ may expose some users' inputs~$\rseq$ at the same time that it leaks very little on others. 
In fact, \mbox{$\e$-in}distinguishability entails bounds on information leakage but the opposite is not generally true~\cite{alvim11,barthe2011information,cuff2016differential}.

Hence, measures like $\e$-indistinguishability are best suited to deal with scenarios with stringent privacy requirements, 
i.e.~where privacy guarantees must hold for every \ot's user and against any possible adversary.
However, 
it is often too expensive or too disruptive to guarantee $\e$-\gls{dp} in many settings, 
e.g.~in the context of location privacy~\cite{bindschaedler2016synthesizing,oya2017geo}. 
By providing an average measure of the protection an \ot\ affords, 
information theoretical measures can better accommodate trade-offs between privacy and utility or cost,
e.g.~by ensuring an acceptable average level of privacy protection, 
even if that means subjecting specific users to greater privacy~loss. 

However, both types of measures, being detached from any particular adversary or attack, are often hard to interpret~\cite{oya2017geo}.
Enforcing differential privacy, it is unclear which value to assign to $\e$ or $\delta$ 
other than $\e =\delta = 0$ when we wish (and it is feasible) to prevent \emph{any} information leakage.
The privacy guarantee ${\e\text{-indistinguishability}}$ provides is \emph{relative}, 
i.e.~it holds regardless of the adversary's background knowledge, yet does not say what an adversary 
armed with sufficient knowledge actually learns.
To determine the \emph{``absolute''} level of privacy,
we must consider a particular adversary with a concrete instance of background knowledge and attack strategy~\cite{shokri2015privacy}.
Information-theoretic measures on the other hand do capture an absolute level of privacy,
i.e.~the number of bits of information leakage; 
however, it is still hard to interpret that number of bits as the risk an \ot's user faces, 
e.g.~the adversary may have enough prior information that 
in spite of how little information the~\ot\ leaks, the user's privacy is breached.
Conversely, by focusing on specific adversaries and attack strategies, \glsreset{aca}\gls{aca}
represents an easier-to-interpret approach to privacy measures, 
arguably better capturing the actual level of \mbox{user privacy~\cite{shokri2011quantifying,shokri2015privacy}}.


\subsection{Attack-centred analysis}

We consider two \glsreset{aca}\gls{aca} measures: \emph{information gain} and \emph{expected estimation error}.

\subsubsection{Information gain}
\label{infoGain}

Information leakage measures the dependence between \ots' inputs and outputs
through the joint probability distribution $\pr{\rseqrv,\eseqrv}$. 
As Chatzikokolakis et al.\ note, we may decompose $\pr{\rseqrv,\eseqrv} = \cp{\eseqrv}{\rseqrv} \cdot \pr{\eseqrv}$,
where $\cp{\eseqrv}{\rseqrv}$ is a parameter from the \ot\ that does not depend on the input probability distribution,
enabling an adversary to, upon having observed an output $\eseq$,
update the probability of input $\rseq$ using Bayes's rule as~\cite{chatzikokolakis2008anonymity}:
\[ \cp{\rseq}{\eseq} = \frac{ \cp{\eseq}{\rseq} \pr{\rseq} }{ \sum_{i} \cp{\eseq}{\rseq_i} \pr{\rseq_i} } \] 

To recover the \emph{`correct'} posterior probability ${\cp{\rseq}{\eseq}}$, however, 
the adversary needs to know the correct prior probability $\pr{\eseq}$,
and this is not necessarily always the case. 
Adversaries may possess incorrect priors that (mis)lead them to estimate a wrong posterior $\prsub{\beta}{\rseq \given \eseq} $~\cite{clarkson2005belief,clarkson2009quantifying}.
Let us denote an adversary's \emph{belief} through the subindex~$\beta$.
When the \ot's output $\eseq$ contradicts an adversary's prior and outputs a sequence~$\eseq$ so that, 
e.g.~$\pr{\eseq \given \rseq_{\prof}} = 0$ while $\prsub{\beta}{\rseq_{\prof}} \rightarrow 1$,
the adversary's uncertainty increases, and yet it gains information to correct her erroneous belief~\cite{clarkson2005belief}.
Conversely, when the \ot's output $\eseq$ strongly supports an erroneous prior, 
the adversary may wrongly recover with high certainty an input $\filseq_{\prof} \neq \rseq_{\prof}$ or, 
by chance, the actual input $\rseq_{\prof}$.\footnote{
These artifacts however tend to disappear with an increasing number of observations,
as the \ot\ leaks more information.}

Moreover, Franz et al.\ warn against adversaries that obtain
side knowledge beyond what the \ot's output leaks,
even if, as Hamadou et al.\ indicate, such side knowledge may also be erroneous~\cite{franz2007attacking,hamadou2010reconciling};
e.g. adversaries may possess background knowledge on a particular \ot's user $\user$,
enabling them to obtain more information about her input $\rseq_\user$ 
than what adversaries that only have access to the population's prior~$\pr{\rseqrv = \rseq}$ 
and obfuscated input~$\eseq$ are able~to.

To incorporate an adversary's (potentially inaccurate) prior and side information into the evaluation, 
previous authors have resorted to \emph{information gain},
defined as the \emph{Kullback-Leibler divergence} ($D_{\text{KL}}$) between an adversary's \emph{belief}
\mbox{---that} includes both prior and side knowledge---
and the actual \ot's input value~$\rseq$ before and after observation~\cite{hamadou2010reconciling}. 
Let~$\abel(\rseq)$ and~$\abel(\rseq|\eseq)$ represent the adversary's beliefs 
that an \ot\ obfuscates some input~$\rseq$
before and after observing the output~$\eseq$,
i.e.~$\abel(\rseq)=\prsub{\beta}{\rseq}$ and $\abel(\rseq|\eseq) = \prsub{\beta}{\rseq \given \eseq}$, respectively.
Let~$\delta(\rseq)$ represent the degenerate probability distribution $\pr{\rseq}$ taking values 
$\pr{\rseq} = 1$ when $\rseq=\rseq_{\prof}$ (the actual input to the~\ot) and $\pr{\rseq} = 0$ otherwise.
We obtain the adversary's information gain~$\gain$~as:
\begin{multline}
 \label{eq:infoGain}
 \gain = \kld{\delta(\rseq)}{\abel(\rseq)} - \kld{\delta(\rseq)}{\abel(\rseq|\eseq)} = \\
 =  \log(\prsub{\beta}{\rseq_{\prof}|\eseq}) - \log(\prsub{\beta}{\rseq_\prof})
\end{multline}
with logarithm base $2$ giving information gain in \emph{bits}.\footnote{
Clarkson et al.\ provide an intuitive interpretation of information gain measured in bits, namely, 
\emph{``$k$ bits of leakage correspond to a $k$-fold doubling of the probability that the attacker
ascribes to reality''}~\cite{clarkson2009quantifying}.
}

Information gain thus provides a measure that is specific to an adversary with 
a particular instance of prior information and side knowledge;
it does not capture an \ot's performance against adversaries with more or less knowledge ---accurate or otherwise. 
While information leakage equals information gain 
if the adversary has a correct prior and no additional side knowledge,
both measures conceptually differ in that information gain measures what the adversary learns and information leakage what the \ot\ leaks. 
The former factors in adversarial beliefs, 
whereas the latter is a function of the inputs and outputs to the channel alone. 

Adversaries with incorrect prior knowledge are in general of lesser concern in \ots' design, 
as wrong prior adversarial \emph{beliefs} hinder adversaries' ability to recover users' data.\footnote{
Hamadou et al.~\cite{hamadou2017quantifying} show that min-conditional entropy is always smaller than
or equal to the min-conditional entropy of an adversary with wrong beliefs, 
i.e.~$\minh(\rseqrv\given\eseqrv) \leq \beliefsup{\minh}(\rseqrv\given\eseqrv)$, 
with $\beliefsup{\minh}$ representing the uncertainty 
of an adversary whose prior \emph{belief} $\prsub{\beta}{\rseq}$ is~incorrect.}
On the other hand, adversaries with accurate side knowledge pose a greater threat than those without,
yet \ots' designers are unlikely to be able to precisely determine the side knowledge adversaries have
or may be able to acquire in the future. 
In general, designing a tool with a particular instance of adversarial beliefs in mind 
renders \ots\ vulnerable to adversaries with beliefs other that what designers account for. 
However, information gain still assists targeted analyses when we wish to ascertain 
the extent to which particular adversaries may breach users' privacy, 
e.g.~as part of an audit or privacy impact assessment.


\subsubsection{Expected estimation error}
\label{par:eee}

\glsreset{lbs}
In the context of their work on privacy-preserving \glspl{lbs}, Shokri et al.\ argue that 
\emph{\guillemotleft[n]either the uncertainty metric nor the inaccuracy metric, however, quantify the privacy of the users.
What matters for a user is whether the attacker finds the correct answer to his
attack, or, alternatively, how close the attacker's output is to the correct answer\guillemotright}~\cite{shokri2011quantifying}.
Shokri et al.'s definitions of \emph{uncertainty} and \emph{inaccuracy} loosely map to
this framework's \emph{information leakage} and \emph{information gain}, respectively.
Their claim echoes our previous observations on the difficulty of interpreting
measures that are detached from a specific adversary or attack
in return for generality: avoiding assumptions about 
adversarial beliefs and attack strategies prevents us from estimating 
how successful the adversary is at retrieving~$\rseq$.
Shokri et al.\ hence propose \emph{expected estimation error},
a probabilistic measure of how \emph{close} the adversary gets to~$\rseq$, 
incorporating a definition of `how bad' it is that the adversary recovers~$\filseq$ instead.
We note that none of the previous measures we have reviewed so far consider
the sequence $\filseq$ the adversary \emph{actually} recovers
or the outcome~$\filprof = \ps(\filseq)$ (as opposed to $\prof = \ps(\rseq)$) that the adversary ultimately obtains;
the measures we have hitherto introduced abstract away from the adversary's function~$\ps$
to provide a measure independent from it. 

We borrow Shokri et al.'s definitions of \emph{expected estimation error} and 
\emph{expected distortion privacy} to define a slight variant of the latter~\cite{shokri2011quantifying,shokri2015privacy}. 
Our goal is to highlight the implications of incorporating the adversary's beliefs, 
the adversary's function~$\ps$ and a measure of adversarial success~$\dist$
to the evaluation of~\ots.\footnote{We refer the reader to Appendix~\ref{appendixEEE} for a detailed explanation of 
how the adaptation of expected distortion privacy introduced in this section differs from Shokri's~\cite{shokri2015privacy}.}
Let $\pr{\prof}$ represent the probability of the adversary's target,
i.e.~$\pr{\prof} = \pr{\ps(\rseq)}$, and $\pr{\eseq \given \prof}$ represent the conditional probability that
the \ot\ outputs~$\eseq$ given that~$\rseq$ maps to~$\prof = \ps(\rseq)$.
Further, let $\pr{\filprof \given \eseq}$ represent the probability distribution that results 
from the attack strategy \emph{and} (estimated) prior knowledge the adversary 
relies on to recover an estimated~$\filprof$ from an observed sequence~$\eseq$.
Lastly, let $\dist(\filprof,\prof)$ represent the distance at which~$\filprof$ and $\prof$ are, 
this distance according to the user's privacy requirements, 
i.e.~the greater $\dist(\filprof,\prof)$ is, the better for user privacy.
We compute \acrlong{eee}, $\eee$, as:
\begin{equation}
 \label{eq:eee}
 \eee = \sum_{\prof} \pr{\prof} \sum_{\eseq} \pr{\eseq\given\prof} \sum_{\filprof} \pr{\filprof \given \eseq} \cdot \dist(\filprof,\prof)  
\end{equation}

Eq.~\ref{eq:eee} incorporates previous measures' assumptions about an \ot's inputs and operation,
and further integrates the adversary's beliefs and attack strategy. 
The first two summations~$\sum_{\prof} \pr{\prof} \sum_{\eseq} \pr{\eseq\given\prof}$,
characterise the channel as information leakage measures do, 
albeit assuming a single input probability distribution,
thus with the limitations and decreased generality of mutual information as opposed to
capacity (which considers all possible probability distributions) 
and indistinguishability (which is independent from the prior $\pr{\rseq}$).
The term~$\sum_{\filprof} \pr{\filprof \given \eseq}$ captures the attack strategy of the adversary
as the probability that the adversary recovers~$\filprof$ 
after updating its prior knowledge with its observation 
of obfuscated input~$\eseq$.
Lastly, the term $\dist(\filprof,\prof)$ incorporates a definition of how $\filprof$ contributes to user privacy, 
namely, it measures how \emph{better} it is for users that the adversary recovers $\filprof$ as opposed to $\prof$~\cite{shokri2015privacy}.

Expected estimation error as defined in Eq.~\ref{eq:eee} departs from previous measures in two significant ways. 
First, it incorporates a distance~$\dist(\filprof,\prof)$ that unambiguously defines 
the advantages of having an adversary recovering $\filprof$ as opposed to~$\prof$.
Note that previous measures, as we have defined them, are \emph{agnostic} to
what it means that an adversary recovers either $\filprof_1$ or $\filprof_2$
as long as they are different from $\prof$, 
i.e.~they do not consider whether it is better or worse that the adversary
recovers $\filprof_1$ or $\filprof_2$ as opposed to $\prof$.
That \emph{``semantic agnosticism''} reinforces their generality at the cost of expressiveness, 
i.e.~an \ot's information leakage~$\mi(\rseqrv;\eseqrv)$ depends on the probability distributions~$\pr{\rseqrv}$, $\pr{\eseqrv}$ and the joint~$\pr{\rseqrv,\eseqrv}$, 
but it is oblivious to the actual values that those random variables take and the metric space we may choose to represent them~on.\footnote{
	Having said that, it is possible to incorporate assumptions about the meaning of $\rseqrv$'s and $\eseqrv$'s values by e.g.~grouping them in bins.}
Second, we note that we could express \gls{eee} 
as a function of the distance~$\dist(\rseq,\filseq)$ instead of~$\dist(\prof,\filprof)$.
With the latter we choose to emphasise the post-processing that ``raw'' data $\rseq$ or~$\filseq$ undergo 
at the hands of the adversary, such as the inference of user attributes which are not explicit
in the observed data~$\rseq$ per se, e.g.~an adversary may collect users' location data $(\rseq)$
to determine their political affiliation ($\prof_1$) and purchasing power ($\prof_2$), 
neither explicit in~$\rseq$, but which the adversary extracts through inference functions~$\ps_1$
and~$\ps_2$, respectively.\footnote{Expressing~\gls{eee} in terms of~$\dist(\rseq,\filseq)$ is a perfectly valid alternative.}
This adds another layer of assumptions about the mapping between~$\rseq$ and~$\prof$ through~$\ps$
which previous measures are also independent (and agnostic)~from.

\begin{figure}
  \centering
	\includegraphics[width=.95\columnwidth]{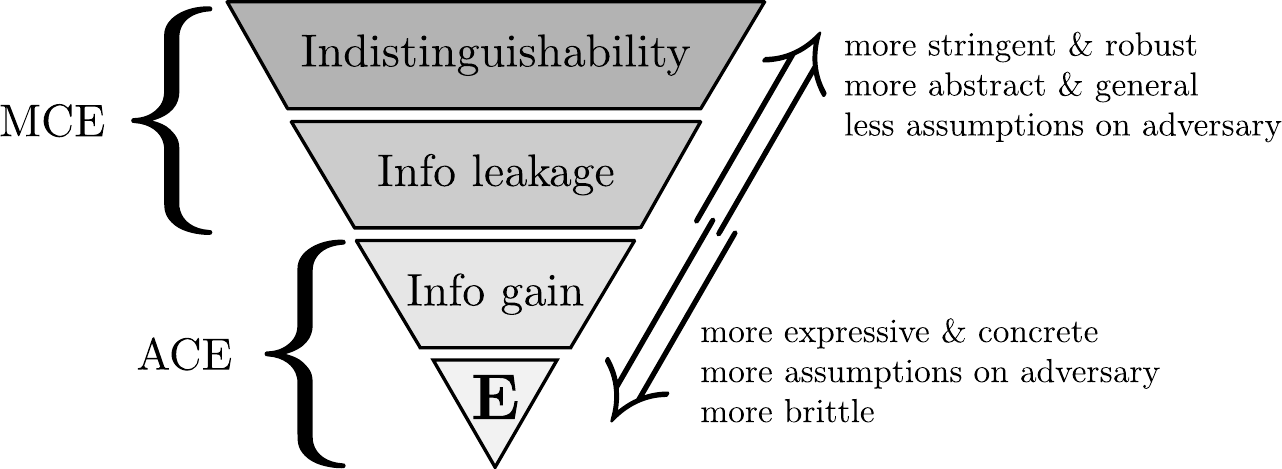}
  \caption[Measure generality given adversarial assumptions]{Hierarchy of measures. Less assumptions confer more generality.}  \label{fig:aMeasures}
\end{figure}

\acrshort{eee} indeed represents the last step in a sequence of measures
from more abstract and general ---less assumptions, less details about the adversary---, 
to more specific and concrete \mbox{---more} assumptions, more details about the adversary. 
We depict this idea in Fig.~\ref{fig:aMeasures}. 
\acrshort{mca}~characterises the extent to which an \ot's protects privacy with 
independence of an adversary's side knowledge and attack strategies.
\acrshort{aca} measures on the other hand incorporate particular details of an adversary: 
information gain considers a particular instance of background knowledge, 
while \acrshort{eee} further considers the particular~$\filprof$
the adversary recovers, thereby implying that not every~$\filprof$ the adversary recovers
is equally \emph{bad} for the user, which in turn depends on a particular adversarial 
post-processing~\mbox{function}.

Yet \acrshort{eee} as an absolute measure of privacy conflates the effect on users' privacy
of the obfuscation mechanism with the adversary's prior and side knowledge as well as its profiling function~$\ps$. 
\acrshort{eee} does not compare users' privacy 
before and after the adversary observes an obfuscated user input~$\eseq$.
An adversary with an exceptionally good prior 
may perform equally well without~$\eseq$, 
highlighting the impossibility of absolute disclosure prevention 
that motivates differential privacy~\cite{dwork2010difficulties}. 
Hence, even if \acrshort{eee} provides a good measure of privacy against a particular adversary, 
it does not properly acknowledge the \ot's ability to prevent privacy~breaches.

\subsection[MCE or ACE?]{\acrshort{mca} or \acrshort{aca}?}
\label{ssc:mcaVSaca}

Disagreements between proponents of different types of measures abound in the literature. 
Andr\'{e}s et al.\ argue that \acrshort{eee} is \emph{``explicitly defined in terms 
	of the attacker's prior knowledge, and is therefore unsuitable for scenarios where the prior is unknown''}~\cite{andres2013geo};
they propose \emph{geo-indistinguishability}, a measure of location privacy based on differential privacy. 
On the other hand, as a relative measure, the $\e$ in $\e$-geo-indistinguishability says very little about 
what the adversary actually knows about the location of a user, as that requires
a specific evaluation that takes an adversary's prior and side knowledge into~consideration.
Similarly, Li et al.~\cite{li2018measuring} argue, in the context of website fingerprint defence evaluation, 
that \emph{``validating [defences] by accuracy alone is flawed [...] when accuracy is low, its corresponding
	information leakage is far from certain''}, and propose mutual information as a measure of information leakage. 
Li et al.'s critique echoes our previous point about the difference between~\gls{mca} and~\gls{aca}. 
The latter conflate the protection an \ot\ affords
with the adversary's prior knowledge and attack strategy. 
Adversaries may successfully infer $\prof$ either due to a weak \ot\ that reveals too much 
or because of robust, prior side information that enables them to obtain~$\prof$
regardless of what little information an \ot\ leaks,
thus underestimating the strong privacy protection the \ot\ would provide in less hopeless scenarios.
Conversely, adversarial failure may result either from a robust \ot\ or from flawed side information 
and a suboptimal attack strategy, the latter leading to the overestimation of an \ot's performance.

Hence, the choice between \gls{mca} and \gls{aca} depends on 
how much generality and stringent a measure of privacy we seek.
\gls{mca} measures provide stronger, more general privacy guarantees
as they do not rest upon specific assumptions on adversarial knowledge, 
i.e.~they are resilient to changes in the attack the adversary deploys and the knowledge it has. 
An \ot's capacity ($\cc$) does not depend on the side knowledge an adversary has; 
neither do differential privacy bounds.
\Gls{mca} measures' generality and independence from contextual details 
makes them suitable as constraints or protection goals in \ot's design,
i.e.~as privacy bounds that \ots\ must generally satisfy. 
However, such generality comes at the expense of expressiveness:
\acrshort{mca} measures do not capture what particular adversaries actually learn or retrieve.
Conversely, by incorporating specific details about the adversary,
\acrshort{aca} provides more expressive measures of privacy loss.
However, this specificity often renders \gls{aca} measures a poor choice when it comes to imposing generic privacy bounds:
because they depend on particularities of the adversary, such as its background knowledge,
they cannot generally hold for \emph{every} adversary, 
e.g.~imposing that an \ot\ guarantees a maximum expected estimation error 
is impossible against adversaries with enough background knowledge, 
who may violate such constraints armed with side knowledge alone 
(i.e.~without exploiting the information the~\ot\ leaks). 
%
Hence, \gls{aca} measures are best suited to determine the threat that particular adversaries 
pose to \ots, to evaluate \ots' deployment in particular contexts, taking into account 
publicly available information that adversaries may exploit or side-channel leakages
that \gls{mca} is oblivious~to.

Lastly, we note that the selection of measures we have provided in this section has allowed us to illustrate 
the differences between the two types of analyses, \emph{mechanism-centred} and \emph{attack-centred},
yet this set of measures is not comprehensive,
with possible variations or combinations of the above also possible, 
depending upon the particular context or \ot\ under evaluation.
Moreover, the conceptual separation between mechanism-centred and attack-centred evaluation
is not always sharply delineated; some measures elude or straddle this binary classification,
e.g.~in the next section we resort to mutual information~$\mi(\rfeat,\profV)$ between a user input~$\rseq$'s features
and the adversary's target~$\prof$ as a measure of information leakage that incorporates assumptions
about the adversarial function~$\ps$. 
In so doing, we lose some of the generality that abstracting away from a particular function $\ps$ confers, 
but we keep some generality by refraining from focusing on a concrete attack strategy or instance of background knowledge. 
In short, the separation between \gls{mca} and~\gls{aca} is not always clear-cut in practice;
rather, it is continuum or spectrum that ranges from the most broad and generic measures 
that avoid most assumptions about the adversary to the most specific measures that focus on very concrete attackers.


\section{Obfuscation and privacy engineering}
\label{s:upoUdoNmore}

The previous section examines how we can 
measure the level of privacy loss~$\priv$ that \ots\ prevent.
In this section, we more generally examine the role of obfuscation in privacy engineering. 
To that end, we explore the relationship between utility and privacy loss,
distinguishing two types of obfuscation, utility-preserving and utility-degrading,
and their respective roles as part of the privacy engineering~toolkit. 

\subsection{Utility-preserving and utility-degrading obfuscation}
\label{upoUdo}

We distinguish between two types of obfuscation: 
\emph{utility-degrading} and \emph{utility-preserving}.
\glsreset{udo}\Gls{udo} reduces users' privacy loss at the expense of utility.
Formally:~\begin{equation}  \osktor(\rseq) = \eseq :  \priv(\eseq) \leq \priv(\rseq) \wedge \util(\eseq) < \util(\rseq)  \end{equation}

	\glsreset{upo}\Gls{upo} modifies $\rseq$ to reduce users' privacy loss while preserving utility. 
	Formally:
	\begin{equation}  \osktor(\rseq) = \eseq :  \priv(\eseq)  \leq \priv(\rseq) \wedge \util(\eseq) = \util(\rseq) \end{equation}

\begin{figure}[t]
	\centering
	\includegraphics[width=0.45\columnwidth]{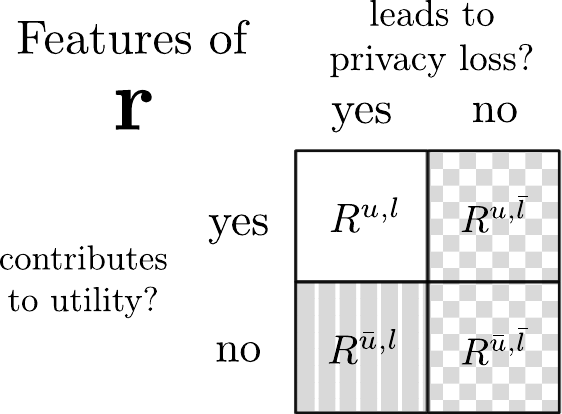}
	\caption{Decomposing~$\rseq$ into features relevant to utility and privacy loss.}
	\label{fig:rfeat}
\end{figure}

Based on the decomposition of a user's input~$\rseq$ into features or random variables $\rfeat$ as we propose in Sect.~\ref{s:model},
we observe the following.
First, to protect a user's privacy, there is no need to obfuscate features~$\rnlfeat$ that provide no information about~$\profV$, 
with $\profV$ the random variable that describes the adversary's target~$\prof$,
i.e.~in terms of information leakage, we refer to features~${\rnlfeat : \mi(\profV ; \rnlfeat) = 0}$.\footnote{
	Note how this formulation implicitly incorporates assumptions about the adversarial function $\ps$ that maps inputs~$\rseq$ 
	to targets~$\prof$. This is not strictly necessary; alternatively we may (rather trivially) refer to features~${\rnlfeat : \mi(\rlfeat ; \rnlfeat) = 0}$.
	Indeed, determining that features~$\rnlfeat$ do not provide information about~$\profV$ 
	depends on a set of assumptions about the underlying \emph{data generation process} and the adversarial function $\ps$~\cite{pufferfish}.
Such assumptions may however not hold in practice.
To achieve the most stringent privacy guarantees, we may assume an arbitrary data generation process
whereby every feature $\rufeat$ provides information about $\prof$, i.e.~$\rseqrv : \forall \rufeat, \mi(\profV \, ; \, \rufeat) > 0$
according to an arbitrary conditional probability distribution~$P(\profV \given \rufeat)$.
}
This holds regardless of whether~$\rnlfeat$ provide utility or not, as these features bring the adversary no closer to obtaining~$\prof$. 
Second, modifying features~$\rnulfeat$ that leak information about~$\prof$
but which functions~$\fui$ and~$\sui$ are oblivious to (so that outcomes~$\ans$ and~$\sans$ do not depend on $\rfeat$) always represents \gls{upo}.
We denote this set of features as $\rnulfeat$ as opposed to the set of features $\rulfeat$ that do intervene in the computation of either $\fui$ or $\sui$ 
\emph{and} leak information about~$\profV$.
Obfuscating~$\rnulfeat$ does not however necessarily lead to decreased privacy loss, 
as $\rnulfeat$~may provide redundant information that the adversary can extract from~$\rulfeat$ alone.
The set of features $\rulfeat$ thus represents 
a major point of contention in obfuscation tools' design, 
as they \emph{may} force a trade-off between utility and privacy. 
We examine the conditions that enable us to escape that trade-off in Sect.~\ref{s:discloseReqs}.
Figure~\ref{fig:rfeat} represents the decomposition of the user input~$\rseq$ into features that intervene in utility and privacy~loss.

\subsection{Factoring in utility: disclosure requirements}
\label{s:discloseReqs}

We observe that if features $\rulfeat$ contribute to personal utility and privacy loss,
there is no inherent trade-off between privacy and \emph{personal} utility~$\util_\fu$.
The system's architecture may impose such a trade-off, but it should be theoretically possible to avoid it with an alternative architecture,
as personal utility alone does not, by definition, require disclosures to untrusted parties.
Conversely, public utility imposes a trade-off with privacy as long as it
depends on features $\rulfeat$ that leak information about~$\profV$,
as producing public utility inherently requires that users disclose information to untrusted,
adversarial parties.

To illustrate how personal utility does not fundamentally require information leakage,
let us consider a video streaming service, such as YouTube,
and a set of users that seek to obtain \emph{only} personal utility from this service, 
i.e.~they want to watch a set of videos without producing any public utility.
As per the current (privacy-unfriendly) system architecture, YouTube requires users
to disclose the list of videos they wish to watch, 
leading to privacy loss because of what the service provider can learn from users' watching patterns. 
There is however no inherent need for users to disclose to YouTube which videos they wish to watch,
as a \acrfull{pir} implementation of this service proves~\cite{gupta2016scalable}.
It is therefore the system architecture that forces users to disclose information 
to meet personal utility requirements, not the functionality itself. 

A trivial universal implementation that illustrates why personal utility 
does not require users to reveal information has the provider sending to users all the data 
and code they need to run the service,
so that they 
do not have to disclose any usage information.
Still, what is theoretically possible may not always be achievable in practice
due to cost, available infrastructure or usability (e.g.~unacceptable delays), among other constraints.\footnote{
A \acrshort{pir} implementation of a streaming service offering a small number of media files (e.g.~in the few thousands) such as Netflix may be feasible~\cite{gupta2016scalable},
but this may not scale well to services such as Google search, indexing \emph{``hundreds of billions of webpages''}
and with stringent latency requirements~\cite{arapakis2014impact,100trPages,taylor2013situation}.
}
Similarly, a service such as Facebook Messenger may require two users to expose their private messages to Facebook itself, 
yet there is no inherent need to do so, as users can exchange messages encrypted end-to-end. 
Moreover, users can even establish a dedicated out-of-band channel that enables them to bypass Facebook altogether, 
which shows that there is nothing inherent in the functionality itself that requires them to disclose their messages
or messaging patterns to Facebook or any other third-party provider, 
as only those two users require access to the messages they exchange.

\begin{figure}
	\centering
	\includegraphics[width=0.75\columnwidth]{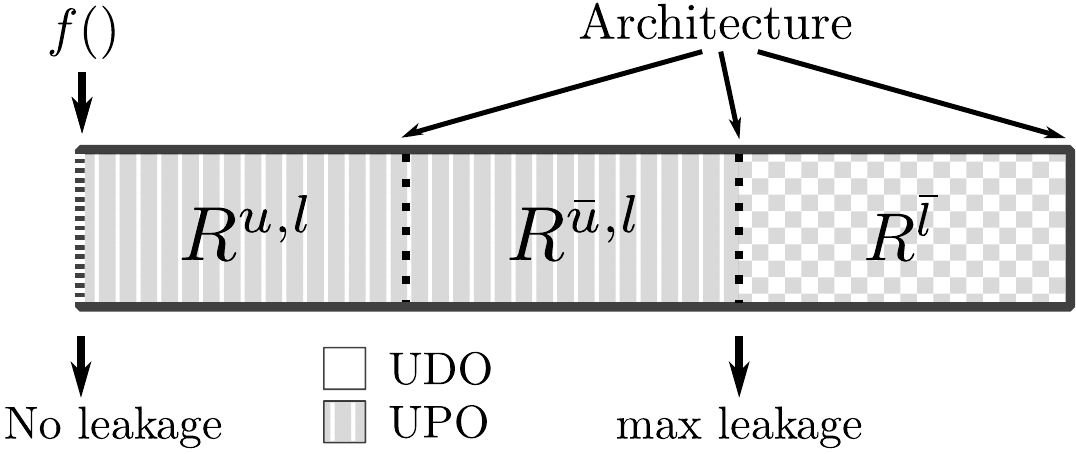}
	\caption[A privacy invasion as information flow]{Personal utility disclosure requirements.\\Personal utility alone should not require~\gls{udo}.}
	\label{fig:persoUtilUPO}
\end{figure}

Hence, whichever personal utility function~$\fu$ users are interested in,
personal utility itself does not inherently require that users disclose information to adversarial parties. 
It is the system's architecture that compels users to reveal \emph{excess information}
for the fulfilment of that functionality. 
Moreover, whereas functionality $\fu$ requires a set of features $\rufeat$ 
to provide the set of outcomes $\aseq$ whence users derive utility, 
the system's architecture may further require that users disclose features $\rnulfeat$
that leak information about~$\profV$ yet are unnecessary for the provision of utility.
Figure~\ref{fig:persoUtilUPO} illustrates this notion:
the horizontal bar represents the amount information that sequence~$\rseq$ leaks to the adversary, 
from zero information (leftmost edge) to maximum information (rightmost edge). 
Personal utility does not require that users provide information to an adversary;
hence, the minimum amount of information about~$\rseq$ that users should reveal is zero, 
which we represent with a tightly dotted line. 
Anything to the right of the tightly dotted line is \emph{excess information},
up to the maximum information leakage limit that features~$\rlfeat$ determine,
as~$\rnlfeat$ do not provide information to the~adversary. 

Utility \emph{producers} on the other hand \emph{always} need to reveal \emph{some} information, 
the rationale being that whoever obtains utility from a utility producer
must have access to at least some function over the producer's data~$\su(\rseq)$. 
Revealing information is therefore \emph{unavoidable}, 
as the intrinsic nature of public utility itself requires it.
Changes to the system architecture can decrease privacy risk by preventing the exposure of information
that is unnecessary for the production of public utility, 
but cannot completely prevent information leakage without destroying all public utility 
---this is, whenever $\rufeat$ provides information about~$\prof$.\footnote{
In other words, we cannot arbitrarily decrease privacy loss without destroying all public utility. 
The rationale follows from the impossibility of absolute disclosure prevention formulated by Dwork and Naor, 
namely, unless we completely randomise $\rseq$ ---in turn resulting in total utility loss---
it is impossible to provide absolute guarantees on what an adversary is able to learn with the information that $\rseq$ discloses~\cite{dwork2010difficulties}.
}

To illustrate how public utility always requires some degree of information leakage,
consider the personalised recommendations on what to watch next that YouTube curates for its users.
Recommendation algorithms typically rely on probabilistic modelling 
to determine relationships between media items that viewers often like,
e.g.~whether viewers that enjoy video~A also often enjoy video~B.
To determine such relationships, recommendation systems require samples 
of the set of videos each person watches.
YouTube's recommendation system relies on a central party (YouTube)
collecting all users' viewing histories, computing a recommendation model,
then offering personalised suggestions to each individual viewer. 
YouTube users thus produce public utility by letting YouTube 
use their viewing histories ($\rseq$) to provide recommendations to others, i.e.~$\sans = \su(\rseq_1, \rseq_2, \ldots, \rseq_N)$.

It is however possible to disclose less information and produce the same level of public utility;
e.g.~we may rely on \gls{pir} \emph{and federated learning} to prevent 
a centralised party (YouTube) from collecting every user's viewing history~\cite{yang2019federated}.
%
With \gls{pir}, each user watches videos from the platform privately, without disclosing her viewing history to anyone. 
With federated learning, users train an individual recommendation model locally ---i.e.~on their own devices--- 
based on their individual viewing histories, which they send to the central party. 
The central party then averages individual models to obtain a global model that 
users download to update their local model and obtain recommendations.
The new system architecture preserves utility \emph{and} improves users' privacy: 
users still benefit from (arguably the same) recommendations but no longer need to disclose the list of videos 
they have watched, only the individual models they compute locally on their devices. 
Indeed, federated learning adds a level of indirection and complexity that prevents 
unsophisticated adversaries from breaching users' privacy;
however, local models still leak information about individual users~\cite{ateniese2015hacking,shokri2017membership}.

To prevent the central party from obtaining individual models,
users may get rid of the central party altogether by running a distributed \gls{mpc} protocol, 
i.e.~instead of computing their individual models locally and sending them to the provider, 
they run a distributed \gls{mpc} protocol among themselves to privately compute a global model
in a way that no participant needs to disclose her individual user model~\cite{damgaard2013practical,archer2018keys}.
Taking this approach to the extreme, rather than disclosing a global model that anyone could download
and try to exploit to extract knowledge about individual media consumption patterns, 
users could simply rely on \gls{mpc} to privately and collectively compute their individual video recommendations,
without disclosing any intervening, intermediary data: not the list of videos they have watched, 
not their individual models, not even a global model. 
This architecture further improves users' privacy without compromising public utility. 

And yet, the resulting recommendations still present biases from each participating utility producer,
leaking information that a knowledgeable adversary could exploit
to extract information about individual users.\footnote{
One particular example of such well-informed adversaries is the classic \emph{``knows-all-records-but-one''} 
or \emph{``has-arbitrary-knowledge''} adversaries that the \emph{differential privacy} literature considers~\cite{kifer2011no},
e.g.~\gls{mpc} does not protect against $n-1$ colluding parties out of $n$ participants.
Let us consider an adversary that observes the recommendation $\su(\rseq_\user,\{\rseq_2, \ldots, \rseq_n\}) = \sans$, 
where $\rseq_\user$ represents the input of the target user 
and the remaining $\rseq_j$ the inputs of the $n-1$ colluding parties.
Knowing $\su$, the colluding parties' $\rseq_j$ and the function output $\ans_{\su}$, 
the adversary could either determine the actual value of input $\rseq_\user$ 
or the set of values that could not have possibly led to $\ans_{\su}$, 
therefore learning \emph{something} about the actual user value $\rseq_\user$
---the exception being that $\su$ is insensitive to the user's input $\rseq_\user$,
in which case the user input is irrelevant, i.e.~public utility does not depend on it.}
However, at this point it is impossible to further reduce the information 
users disclose without a drop in utility: we have redesigned the system to 
only disclose what is strictly necessary for the production of public utility, 
namely, recommendations themselves. Any intervening data in such computation is not explicitly exposed,
but the production of public utility inherently requires that we disclose \emph{some} information, 
this is, those bits of information from $\rseq$ that contribute to produce those recommendations.
Hence, users must resort to some form of \gls{udo}, e.g.~randomised response,
to further minimise the privacy risk they incur by contributing to the production of public utility~\mbox{\cite{beimel2008distributed,kairouz2016differentially}}. 

\begin{figure}
	\centering
	\includegraphics[width=0.75\columnwidth]{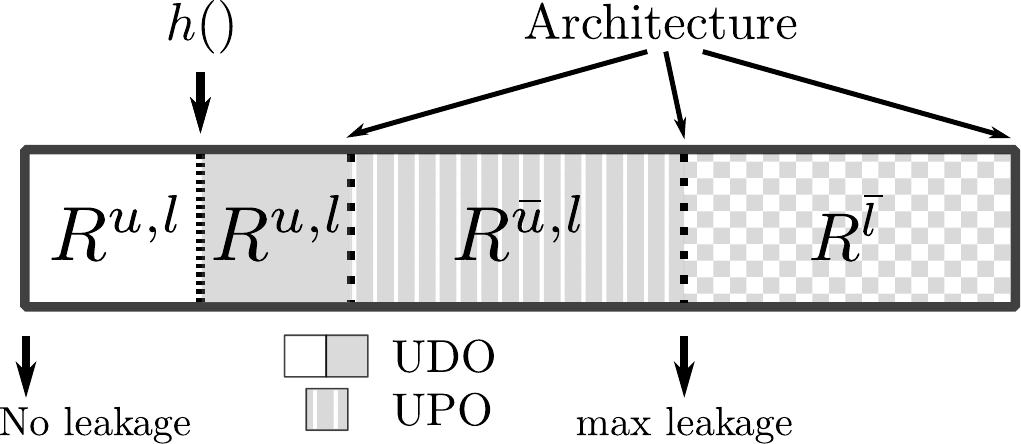}
	\caption[]{Interplay between public utility exposure requirements, a system's architecture and obfuscation.}
	\label{fig:publicUtilityObfs}
\end{figure}

This succession of system (re)designs enable us to see that there are two types of disclosure requirements:
those that \emph{public} utility imposes, and those that the system architecture imposes. 
The system architecture may require that we expose more information than what is 
actually necessary to reveal, i.e.~disclose~$\rseq$ or some function thereof
as opposed to the minimum amount of information that public utility requires, 
i.e.~the outcome $\sans = \su(\rseq)$. 
Figure~\ref{fig:publicUtilityObfs} depicts this idea:
The system architecture requires users to expose,
from left to right, features $\rulfeat$ that contribute to public utility and leak information to the adversary (white and grey backgrounds),
features $\rnulfeat$ that \emph{do not} contribute to public utility but still leak information to the adversary (striped background) 
and features $\rnlfeat$ that may or may not contribute to public utility, but they do not leak information to the adversary (chequered background).
The tightly dotted line where $\su()$ points to marks the minimum amount of information 
the user must disclose without public utility loss; the portion of the bar to the right of this line
represents \emph{excess} information that the architecture ---unnecessarily--- leaks.

The disclosure requirements that either personal or social utility impose 
thus have key implications for privacy engineering.
Since personal utility does not inherently require users' data disclosure, 
systems that do require utility consumers to disclose their data to adversarial parties
reveal an unequivocal culprit:
the system does not satisfy privacy-by-design requirements, 
requiring that users reveal more information than what is strictly necessary, 
be it for reasons of cost, lack of expertise or intentional data siphoning~\cite{gurses2015engineering}.
Hence, utility \emph{consumers} should not be forced to \emph{unilaterally} resort to~\ots\
to protect their privacy (q.v.~Figure~\ref{f:modelWithObfs})
even if obfuscation represents a valuable tool for users to contest privacy-invasive system architectures~\cite{brunton2015obfuscation}.
Instead, as Figure~\ref{fig:persoUtilUPO} suggests, 
system designers should ideally \emph{embbed} \gls{upo} or alternative privacy technologies
in the system design,\footnote{
As illustrated earlier in this section,
system designers may resort to utility-preserving privacy technologies 
that do not rely on obfuscation, e.g.~through the use of cryptographic primitives
such as \gls{pir}.}
e.g.~\emph{embedding} in the system the user-deployed \ot\ that Fig.~\ref{f:modelWithObfs} depicts 
to protect utility consumers' privacy \emph{by design}.
As we argue in the next section, exclusively relying on \gls{upo} 
to protect utility consumer's privacy is indeed theoretically, 
conceptually possible through the use of \emph{chaff}.
\gls{udo} does however represent an alternative in those situations where the system architecture 
or high costs prevent the deployment of \gls{upo} in practice, 
even if the trade-off between utility and privacy that \gls{udo} imposes 
is not fundamentally necessary.

Conversely, since social utility mandates users' data disclosure (or a function thereof),
once we expose nothing but the strictly necessary information about $\rseq$
that the production of public utility requires, 
there is no workaround to the trade-off between public utility and privacy.
At this point,  we \emph{must} resort to (utility-degrading) obfuscation to remove information that 
both intervenes in the production of public utility and contributes to privacy loss, 
meaning that once we reach the point at which the system exposes
nothing but the minimum information required to fully guarantee public utility, 
\acrshort{udo} represents {the only solution} to further decrease users' privacy loss.
Returning to the YouTube example above, when users engage in a distributed \gls{mpc} protocol 
to privately learn their recommendations, 
they expose the minimum amount of information to preserve public utility;
from that point onwards, participating users can only resort to some form of input randomisation 
(\`a la differential privacy) 
to further diminish the privacy loss they incur in the production of public utility.
Such input randomisation leads to utility-loss and thus represents \gls{udo}.
Hence, obfuscation is key to enable individuals to balance
the privacy loss they incur in the production of public utility.
System designers may thus incorporate obfuscation on behalf of utility producers 
(see~Fig.~\ref{f:modelWithObfs2}), provide support for user-generated obfuscation 
(as in~Fig.~\ref{f:modelWithObfs}) or both (Fig.~\ref{f:modelWithObfs3}).

A system's architecture may however also require that users expose more information 
than what is strictly necessary for the production of public utility,
e.g.~YouTube requires that users disclose their viewing histories to serve recommendations to others. 
In such cases, we can deploy \gls{upo} to remove information about features $\runlfeat$ that do not intervene in the production of public utility;
however, features $\rulfeat$ require the deployment of \gls{udo} even if that set of features still leaks more information than what is strictly necessary.
Returning to the YouTube example above, 
the interventions we propose (i.e.~computing the global model using federated learning and \gls{mpc}) 
prevent the disclosure of \emph{some} excess information while preserving all public utility,
but still leak more than strictly necessary, forcing users that do not wish to expose as much information about $\rseq$ to resort to \gls{udo}.
Figure~\ref{fig:publicUtilityObfs} represents this interplay between obfuscation, public utility and a system's architecture:
the tightly dotted vertical line represents the minimum amount of information about $\rseq$ that users must expose to preserve all public utility.
The white area to the left of that line represents the range where a trade-off between privacy and utility is unavoidable: further decreasing privacy loss entails diminished public utility. 
Navigating such trade-off necessarily requires the deployment of \gls{udo}.
The grey area to the right of that line represents excess information embedded in features $\rulfeat$ that the system's architecture exposes
to fulfil public utility requirements, thereby forcing the deployment of \gls{udo} to reduce privacy loss even if such a trade-off is not strictly necessary.



\section{Utility-preserving obfuscation and chaff}
\label{upoNchaff}

\begin{figure*}
	\centering
	\includegraphics[width=0.75\textwidth]{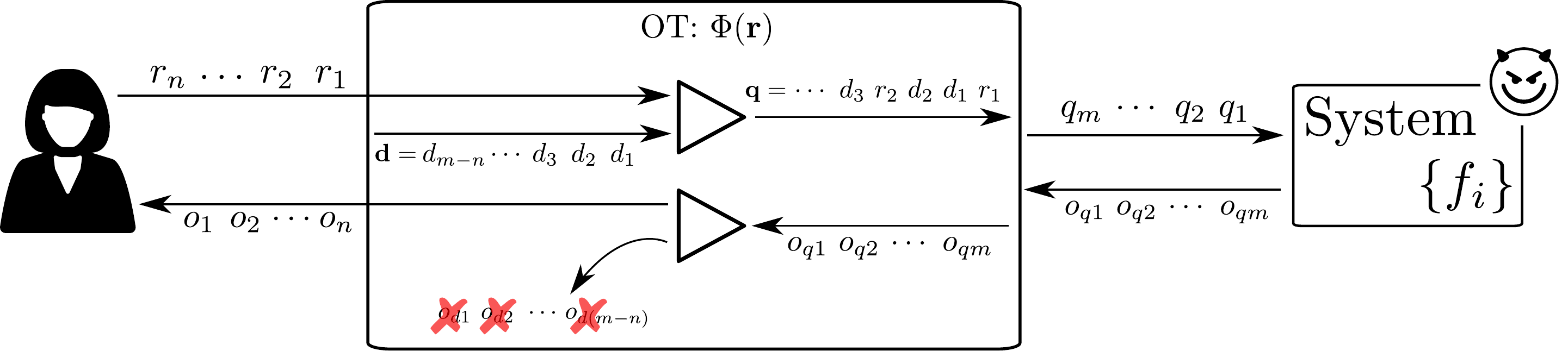}
	\caption{Operation of a chaff-based obfuscation tool}
	\label{fig:chaffModel}
\end{figure*}

Utility-preserving obfuscation (\acrshort{upo}) requires that users' inputs 
are either left intact or replaced by inputs that result in identical or equivalent utility 
(i.e.~that result in the same outputs~$\ans$ and~$\sans$).
Interleaving the user's genuine input~$\rseq$ with a sequence~$\dseq$ of \emph{chaff}, 
this is, dummy or fake user activity, 
pollutes an adversary's observation but does not modify the user's individual activities~$\real_i$.
Under the assumption that chaff is \emph{indistinguishable} from the user's real activity
 ---the key requirement for any chaff-based obfuscation strategy to work---,
the adversary's observes a \emph{polluted} set of input data~$\eseq$
that should ideally reveal less or no information about $\rseq$, thereby reducing privacy loss~\cite{ebalsaPhdThesis}.
Moreover, if the generation of chaff meets certain conditions (which we review below) 
and the \ot\ can filter out responses $\ans_{\dummy}$,
the user receives the original output $\aseq$, hence the same utility~\cite{obpws}. 
Chaff can thus prevent privacy loss \emph{and} preserve personal utility, enabling \gls{upo}.
Figure~\ref{fig:chaffModel} depicts the system model featuring a chaff-based~\ot.
Below we illustrate through three different scenarios how to perform \gls{upo} with chaff.

Let us return to the previous video streaming use case (e.g.~YouTube).
First, let us consider the service offers no personalisation,
i.e.~the set of results $\ans = \fu_{0}(\real)$ (a list of videos) any search query $\real_i$ leads to  
does not depend on previous queries by \emph{any user}.
In this scenario, an \ot\ can filter out responses to dummy queries $\ans_{\dummy}$
that the service returns to the user ensuring that responses $\ans_{\real}$ 
remain the same as if no dummies were added, preserving personal utility.
However, the adversary observes a sequence of real and dummy queries~$\eseq$
which ideally prevents it from retrieving the original~$\rseq$.

Second, let us consider the service offers individual personalisation, 
i.e.~each query~$\real_i$ depends on previous queries \emph{by the same user}.
The search function~$\fu_{1}$ takes a query~$\real_i$ and the sequence~$\rseq = [r_1,\dots,r_{i}]$ of previous queries (or subset thereof)
to produce a list of videos $\ans$, i.e.~$\ans = \fu_{1}(\real_i, \rseq)$. 
In this scenario, an \ot\ cannot simply generate dummy queries and filter 
the corresponding dummy responses to preserve utility, 
as $\fu_{1}$ would feed on an obfuscated input $\eseq$ that contains dummy queries, disrupting personalisation, 
i.e.~$\fu_{1}(r,\eseq) \neq \fu_{1}(r,\rseq) \Rightarrow \util_{\fu}(\eseq) < \util_{\fu}(\rseq)$.\footnote{
	Note that this implicitly assumes search results become less useful 
	if the personalisation function feeds on dummy queries $\dummy_i$ unrelated to the user's search queries $\rseq$. 
}
To overcome this issue, the \ot\ generates chaff by creating dummies of 
\emph{the least reducible unit of data} that the function the user obtains utility from (i.e.~$\fu_{1}$) \emph{feeds} on, 
namely, dummy units of the entire user input $\rseq$.
To~do so, the~\ot\ resorts to Sybil accounts:
the \ot\ generates dummy user accounts on which it sends dummy queries, 
but keeps the genuine user account free of chaff.
The honest-but-curious service provider returns personalised search results
for the real user account and each of the dummy accounts independently, 
thus preserving utility for the user. 
In terms of privacy, assuming indistinguishability between the real user account 
and the Sybils, the adversary cannot distinguish the real input $\rseq$,
which ideally results in reduced privacy loss, i.e.~$\priv(\eseq) < \priv(\rseq)$.\footnote{
One may protest that, first, this configuration does not protect the user against practices like profiling
because, since the user's actual account is free from chaff,
the adversary can still subject the user to unfair or discriminative targeted advertising.
In fact, and second, the adversary may be unaware of the existence of Sybil accounts
and fail to link them to the \ot's user.
We recall that the model in Sect.~\ref{s:advmodel} specifies a \emph{strategic} adversary whose goal is to recover $\prof = \ps(\rseq)$,
which means that it \emph{knows} the \ot\ generates Sybil accounts 
and the consequences derived from the recovery of $\prof$ or $\filprof$
(such as targeted advertising) are out of scope.
Note that, even in the previous scenario ($\fu_{0}$),
the \ot\ cannot prevent profiling based on a filtered~$\filprof$ 
that ultimately leads to worse outcomes on the user.
This further highlights the protective goal that obfuscation plays in this setting:
it attempts to prevent the adversary from obtaining (additional) knowledge, 
rather than prevent subsequent processing that may take place regardless.
Addressing this issue may require a departure from privacy technology 
to solutions that focus on the impact of algorithmic decision-making, 
such as protective optimisation technologies~\cite{pots,kulynych2020pots}.
}

Third, let us consider a service with \emph{collective} personalisation, 
this is, each query depends on other users' previous queries.
The search function $\fu_{2}$ takes a query $\real_{\user}$ from user $\user$ 
and previous queries~$\rseq$ by the querying user and other users,
 i.e.~$\ans = \fu_{2}(\real_{\user}, \rseq_{\user}, \rseq_{1}, \ldots,\rseq_{N-1})$, 
with~$N$ the user population size.   
We further consider the user to exclusively be a utility consumer, 
i.e.~she does not wish to produce public utility and disregards
obfuscation's impact on personalised results for others. 
Still, an \ot\ that relies on Sybil accounts like in the previous scenario is bound to disrupt global personalisation
for everyone, \emph{including} the \ot~user:
the service provider may choose to entirely disregard the user input (either real or dummy) 
to prevent polluting the collective personalisation model or, 
instead, select as input the user account it identifies as real (being potentially a Sybil), among other strategies.
As a result, the user's input is either absent or incorrect, 
affecting collective personalisation for everyone, 
changing the search results that other users' obtain for their queries, 
thus further altering the feedback they provide to the personalisation function.
In this particular scenario, to deploy \gls{upo} with chaff,
the service provider would need to generate independent collective personalisation 
for each of the user's Sybil accounts, which would in turn require the cooperation
of every other user, by providing feedback to the collective model created for every Sybil account. 
Needless to say, this represents a ridiculously expensive and unrealistic solution.
However, it still shows how chaff, \emph{theoretically}, enables \gls{upo}.
Finally, this third scenario further showcases how, to benefit from collective personalisation,
 the user depends on everyone else's willingness to produce public utility:
if everyone selfishly wished to enjoy collective personalisation without revealing data about themselves,
collective personalisation would not be possible in the first place.
Someone must give away some or their privacy to produce utility for everyone else. 

The fact that the generation of chaff or dummies is often prohibitively costly
points in fact to a different trade-off, namely, not between utility and privacy, 
but between privacy and \emph{cost}.
Chaff-based \ot\ or non-obfuscation-based solutions such as \gls{pir} or \gls{mpc} 
preserve users' utility yet incur additional (and sometimes prohibitive) 
costs for either users or service providers, or both.


\section{Conclusion}

Obfuscation represents an essential set of techniques in the privacy engineering toolkit. 
Yet a holistic understanding of the role that obfuscation, in general, plays in the engineering of privacy 
has hitherto been elusive.
Furthermore, the multiplicity of contexts in which obfuscation technology has been applied
has led to a vast array of privacy measures to evaluate the protection that obfuscation affords. 
This fragmented landscape of obfuscation technologies and piecemeal evaluation methods 
means that we also lack a cohesive, comprehensive understanding of the conditions and assumptions 
under which any particular measure suitably assists the evaluation of obfuscation technologies. 

In this paper we take a step back from each of the specific settings
in which obfuscation is applied to provide a set of general observations
that apply to the broader family of obfuscation technologies as a whole. 
The core contribution of this paper is a conceptual framework whose aim is to enable a better understanding 
of the role that obfuscation technology plays in the engineering of privacy. 
To that end, we have proposed an abstract model of obfuscation, 
bringing together a multiplicity of obfuscation methods under the same analytical umbrella, 
whether they are deployed by users, trusted curators or service providers, or both.
Moreover, to advance a better grasp of how to evaluate obfuscation technologies, 
we have proposed a conceptual separation between two main types of evaluation: mechanism-centred and attack-centred. 
Through this conceptualisation, we have unpacked underlying assumptions to the measures that fall within either approach and shed further light on the conditions that call for choosing one measure or another, 
e.g.~mechanism-centred measures are best suited to impose privacy bounds 
while attack-centred measures enable us to measure the privacy loss 
that particular, concrete adversaries subject users to.

We have introduced the concepts of \emph{personal} and \emph{public utility}
as well as, correspondingly, \emph{utility consumer} and \emph{producer}.
We have distinguished two types of obfuscation: utility-preserving (\gls{upo}) and utility-degrading (\gls{udo}),
and observed that the production of public utility \emph{requires} 
the deployment of \gls{udo} to arbitrarily reduce privacy loss, 
whereas personal utility alone does not impose a trade-off between utility and privacy, 
thereby \emph{theoretically} enabling the deployment of \gls{upo} 
or alternative utility-preserving privacy technologies that do not rely on obfuscation
---with \gls{udo} being forced by extrinsic reasons such as cost and adversarial service providers. 
Lastly, we have illustrated how to perform \gls{upo} through \emph{chaff}, 
acknowledging that escaping the trade-off between utility and privacy 
invokes a different trade-off between privacy and cost.

This work opens up a multitude of future research avenues:
The separation between mechanism-centred and attack-centred evaluation
can contribute to spawn more systematic approaches to the evaluation of obfuscation technologies,
providing a framework to compare designs and evaluations that do not share 
a common set of adversarial assumptions.
Distinguishing between personal and public utility has an impact on privacy-by-design principles and policymaking, 
e.g.~should users be able to completely opt-out from public utility production? 
In other words, should users be able to participate in a system exclusively as utility consumers?
Furthermore, fully addressing the trade-off between public utility production and privacy 
requires utility-\emph{degrading} obfuscation, thus raising questions about the impact
that obfuscated data has on subsequent algorithmic decision making. 
How does the privacy protection through obfuscation interact with 
other goals and requirements, such as algorithmic fairness~\cite{kulynych2020pots}?
Future work should also advance a better understanding of \gls{upo} methods beyond chaff, 
as well as to explore further the underlying rationale to different obfuscation technologies.

\section*{Acknowledgments}

The author would like to thank Ward Beullens and Charlotte Bonte for answering questions about notation and cryptographic obfuscation. 
This work was supported by CyberSecurity Research Flanders with reference number VR20192203.

\bibliographystyle{IEEEtran}
\bibliography{onObfs,obfsSoK,chaff}

\appendix


\subsection{On alternative measures of information leakage}
\label{appendixMI}

Both mutual information and capacity are based on Shannon's entropy which,
according to Cachin's interpretation, 
measures information as \emph{``the average number of binary questions
an adversary needs to ask about $\rseq$ to determine its value''}~\cite{cachin1997entropy}. 
Mutual information thus measures information leakage in terms of 
the average number of binary questions one no longer has to ask
after observing~$\eseq$, i.e.~reducing uncertainty about~$\rseq$.
However, this may not always be the most meaningful measure of information leakage. 

Among other authors, Smith has prominently questioned the research community's over-reliance in Shannon's entropy as a one-size-fits-all measure of uncertainty~\cite{smith2011quantifying}.
Smith shows that Shannon's entropy underestimates the \emph{vulnerability} of 
the most likely~$\rseq$~\cite{smith2011quantifying}, 
i.e.~if we are mostly concerned about the information an~\ot\ leaks about 
the most likely input~$\rseq$.
Pliam also shows that there is an unbounded gap between Shannon's entropy 
and \emph{marginal guesswork}, related to the minimum number of searches a brute-force attacker 
must perform to attain a certain probability of success~\cite{pliam2000incomparability}.
This means that the higher the probability $\max(\pr{\rseqrv=\rseq})$,
the more likely the adversary guesses it correctly \emph{regardless of obfuscation};
yet Shannon's entropy may remain arbitrarily high as long as there are sufficient alternative 
sequences $\rseq'$ with greater than zero (even if negligible) probability.
These critiques illustrate that there are multiple ways of measuring 
the amount of information an \ot\ leaks, e.g.~min-entropy-based measures 
such as min-entropy leakage and min-capacity
capture better the information that \ots\ leak about the most likely input~$\rseq$, 
whereas measures derived from marginal guesswork may better capture an \ot's leakage 
of information in a setting where multiple guesses are allowed.
More recently, the notion of $g$-leakage provides a generalisation of min-entropy 
that uses \emph{gain functions} to capture a variety of adversarial attack~strategies~\cite{m2012measuring}.

These information leakage variants thus capture 
different notions of the type of information an~\ot\ should \emph{not} leak, 
e.g.~min-leakage measures the amount of information an~\ot\ leaks about 
the most likely input~$\text{arg\,max}_{\rseq}(\pr{\rseq})$, 
essentially disregarding the amount of information the~\ot\ leaks about less likely inputs.
Hence, these notions implicitly incorporate assumptions about the kind of information 
that may be useful to an adversary given a general attack strategy:
be it to recover a single~$\filprof$ with the highest probability of being $\prof$
or select the top $k$ most likely $\filprof$, 
carrying on under the assumption that any of those inputs may be the actual~$\prof$ 
---e.g.~to lessen the probability of focusing on a single~$\filprof$ that does not~match~$\prof$.
Analysts and designers must therefore carefully select the measure 
that best represents the kind of information the~\ot should not leak.

\subsection{On Shokri's \emph{expected distortion privacy}}
\label{appendixEEE}

Shokri's definition of \emph{expected distortion privacy} 
uses $\pi(\rseq)$ instead of $\pr{\rseq}$, 
with $\pi(\rseq)$ representing $\pr{\rseq}$ 
as given by prior, publicly available information~\cite{shokri2015privacy}. 
Shokri notes that $\pi(\rseq)$ does not however denote an adversary's auxiliary knowledge, 
which is generally unknown, i.e.~$\pi(\rseq)$ represents an estimation 
of what the adversary knows based on publicly available information,
not what the adversary \emph{really} knows (e.g.~through other information channels that neither we know nor control).
Using $\pi(\rseq)$ instead of $\pr{\rseq}$ is consistent with Shokri's goal, namely, 
to design an obfuscation mechanism that \emph{optimises} the obfuscation strategy 
based on previous disclosures assuming an informed adversary that uses an optimal attack strategy.
Conversely, we consider an adversary with potentially incorrect prior knowledge
and suboptimal attack strategy. 
Hence, we use $\pr{\rseq}$ to average over the actual probability of inputs $\rseq$
and incorporate the adversary's prior knowledge (which, as Shokri, we acknowledge is an estimation, 
not the actual knowledge) in the term that Shokri reserves for the inference attack alone, 
i.e.~$\pr{\filseq \given \eseq}$.
These changes enable us to measure the expected estimation error of adversaries with incorrect priors.

\end{document}